\documentclass[final]{ifacmtg}
\usepackage{graphicx,latexsym,blackboard}
\def\A{{\cal A}}

\def\C{{\cal C}}
\def\D{{\cal D}}

\def\J{{\cal J}}
\def\H{{\cal H}}
\def\L{{\cal L}}

\def\eps{\epsilon}
\def\boldsymbol#1{\mathbf{#1}}
\def\op#1{\hat{#1}}
\def\op#1{#1}
\def\ket#1{| #1 \rangle}

\def\bra#1{\langle #1 |}

\def\Tr{\mathop{\rm Tr}\nolimits}

\newif\ifpdflatex\pdflatextrue
\makeatletter\@ifundefined{pdfoutput}{\pdflatexfalse}\makeatother
\def\myincludegraphics[#1]#2#3{%
    \ifpdflatex \includegraphics[#1]{#2}
    \else       \includegraphics[#1]{#3}
    \fi}
\begin{document}
\begin{frontmatter}\vspace{-17ex}
\title{Hamiltonian Engineering for Quantum Systems}
\author{Sonia G Schirmer}\vspace{-2ex}
\address{Dept of Applied Math \& Theoretical Physics,
         University of Cambridge,  Cambridge, CB3 0WA, United Kingdom}\vspace{-2ex}
\begin{abstract}
We describe different strategies for using a semi-classical controller to engineer 
quantum Hamiltonians to solve control problems such as quantum state or process 
engineering and optimization of observables. \textit{Copyright \copyright\ 2006 IFAC}.
\end{abstract}
\end{frontmatter}

\section{Introduction}

Extending control to the quantum domain, i.e., to physical systems whose behavior
is not governed by classical laws but dominated by quantum effects, has become an
important area of research recently.  It is also an essential prerequisite for the
development of novel technologies such as quantum information processing, as well
as new applications in quantum optics, quantum electronics, or quantum chemistry.
Choreographing the behavior of interacting quantum particles is a rather difficult
task in general, for a variety of reasons, including the destructive effects of
uncontrollable interactions with the environment and measurement backaction, both
of which lead to loss of coherence, for instance.  Yet, many problems in quantum 
optics, quantum electronics, atomic physics, molecular chemistry, and quantum 
computing can be reduced to quantum state or process engineering, and optimization
of observables, which can be solved using Hamiltonian engineering techniques.  In
this paper we outline and compare different strategies for accomplishing this in 
various quantum settings.

\section{Control System Model}

\begin{figure}
\center\myincludegraphics[width=2.5in]{figures/png/CSModel2.png}{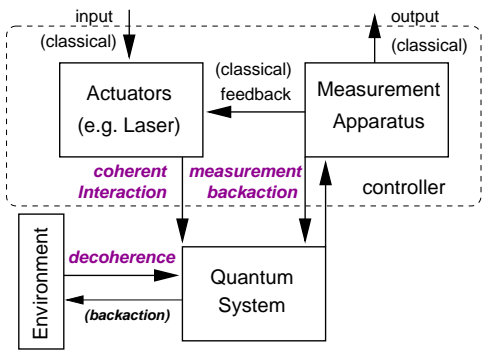}
\caption{Semi-classical Quantum Control Model} \label{fig1}
\end{figure}

A control system must obviously consist of a system to be controlled and a controller.  
In quantum control the former is always quantum-mechanical.  The latter can be either
quantum or classical, and it may seem natural to choose another quantum system as the
controller.  Indeed, this is useful for some applications, for instance in quantum 
optics~[\cite{IEEETAC48p2107}].  The main challenge in general though, and the focus
of this paper, is control at the interface between the quantum world and the classical
world we experience, i.e., control of systems that obey the laws of quantum physics 
using semi-classical sensors and actuators that interact with the quantum system but
accept classical input (such as different settings of the classical control switches 
of the laboratory equipment) and return classical information (See Fig.~\ref{fig1}).  

This model is applicable to many quantum control settings.  For example, the quantum
system could be an ensemble of molecules involved in a chemical reaction, subject to 
laser pulses produced by an actuator consisting of a laser source and pulse shaping 
equipment, and a detector that might consist of a mass spectrometer to identify the 
reaction products.  It could be a solid-state system such as an ensemble of quantum 
dots representing qubits, with actuators and sensors consisting of control electrodes
and single-electron transistors, respectively.  It could be an ensemble of molecules
with nuclear spins subject to actuators generating magnetic and radio-frequency fields,
and sensors that detect the magnetization of the sample, etc.  

\section{System Dynamics}

In the setting defined above the state of the system can be represented either by 
a Hilbert space vector $\ket{\Psi(t)}$, or more generally, a density operator 
$\op{\rho}(t)$, i.e., a positive trace-one operator acting on the system's Hilbert 
space $\H$.  Although the full Hilbert spaces of most quantum systems are hardly 
finite or separable---even the Hilbert space of the hydrogen atom is not separable 
if the full continuum of ionized states is included, we are usually only interested 
in controlling the dynamics of a finite-dimensional subspace of the system's full 
Hilbert space (e.g., a subset of bound states of an atom), and full control of an 
infinite-dimensional system is generally not possible in any case [see 
e.g.~\cite{IEEE39CDC2803}].  We will therefore assume in this paper that the Hilbert 
spaces of interest are finite-dimensional, and thus all operators have matrix 
representations, etc.

Neglecting decoherence and the effect of measurements for a moment, the evolution 
of the quantum system to be controlled is governed by the Schrodinger equation
\begin{equation} \label{eq:SE}
  \frac{d}{d t} \ket{\Psi(t)} = -\frac{i}{\hbar} \op{H}[\vec{f}(t)] \ket{\Psi(t)},
\end{equation}
or the (equivalent) quantum Liouville equation
\begin{equation} \label{eq:LE}
 \frac{d}{dt} \op{\rho}(t) = -\frac{i}{\hbar}\left[\op{H}[\vec{f}(t)],\op{\rho}(t)\right],
\end{equation}
where $[A,B]=AB-BA$ is the commutator.  $\hbar=h/2\pi$, where $h$ is the Planck 
constant, but we will often choose units such that $\hbar=1$.  Thus, the dynamics 
is determined by the Hamiltonian $\op{H}[\vec{f}(t)]$, which is an operator acting
on the system's Hilbert space $\H$, and depends on a set of (classical) control 
fields $\vec{f}(t)=\left(f_1(t),\ldots,f_M(t)\right)$ 
produced by the actuators.  

The main difference between Eqs (\ref{eq:SE}) and (\ref{eq:LE}) is that the former
applies only to pure-state systems, while the latter applies equally to pure and 
mixed-state systems (quantum ensembles).  Moreover, unlike the Schrodinger equation,
the quantum Liouville equation can be generalized for systems subject to decoherence 
or weak measurements by adding (non-Hamiltonian) superoperators [operators acting on
Hilbert space operators such as $\op{\rho}(t)$] to account for the contributions of 
measurements and dissipation to the dynamics of the system
\begin{equation} \label{eq:LE2}
  \dot{\op{\rho}}(t) = \L_H[\op{\rho}(t)]+\L_M[\op{\rho}(t)]+\L_D[\op{\rho}(t)], 
\end{equation}
where $\L_H[\op{\rho}(t)]=-\frac{i}{\hbar}\left[\op{H}[\vec{f}(t)],\op{\rho}(t)\right]$.
The main difference between $\L_M$ and $\L_D$ is conceptual: the former depends on 
the measurements performed, i.e., the configuration of the sensors, which we can 
control, the latter on usually uncontrollable interactions of the system with the 
environment.


Due to the physically prescribed evolution equations for quantum systems, quantum
control---at least in the classical controller model described---is fundamentally
non-linear, even in the case of open-loop control.  In the special case when the 
system's interaction with both sensors and the environment is negligible (which is
obviously only possible for open-loop control), the system's dynamics is completely
determined by a Hamiltonian operator $\op{H}[\vec{f}(t)]$, and the main task of 
quantum control is to find effective ways to engineer this Hamiltonian to achieve 
a desired objective.  For simplicity we will restrict our attention here mostly to 
Hamiltonian engineering in this case, although many of the techniques are still
useful in the more there general case when the system is subject to measurements 
and/or uncontrollable interactions with the environment [Eq.~(\ref{eq:LE2})].

\section{Quantum Control Objectives}

Although the objectives vary depending on the application, most problems in quantum 
control can be reduced to quantum process engineering, quantum state engineering 
or optimization of observables.  The first, quantum process engineering, involves 
finding a Hamiltonian $\op{H}[\vec{f}(t)]$ such that
\begin{equation}
  \exp_+\left[ -i\int_{t_0}^{t_F} \op{H}[\vec{f}(t)] \,dt \right] = \op{U}, 
\vspace{-2ex}
\end{equation}
where $\op{U}$ is a desired target process, and the subscript $+$ indicates that 
the exponential must be interpreted as a time-ordered exponential with positive 
time-ordering due to the time-dependent nature of the Hamiltonian.  This problem 
is of particular interest in quantum computing, where the target processes are 
quantum logic gates.  In general, not every process can be implemented for a given
system, and the set of reachable processes is determined by the dynamical Lie group 
associated with the set of Hamiltonians $\{\op{H}[\vec{f}(t)]:\vec{f}(t)\in \A\}$, 
where $\A$ is the set of admissible controls that can be produced by the actuators
and satisfy physical constraints, e.g., on the field strength, etc.  For example, 
for a Hamiltonian system a necessary (minimum) requirement is that $\op{U}$ be a 
\emph{unitary} operator (acting on the system's Hilbert space), and other 
constraints may restrict the set of reachable operators further.  However, if the
target process is reachable then there are usually many Hamiltonians, which will 
give rise to the same process.  

The second problem, quantum state engineering, requires finding a Hamiltonian 
$\op{H}[\vec{f}(t)]$ such that (a) $\ket{\Psi(t_F)}=\ket{\Psi_1}$, $\ket{\Psi(t_0)}
=\ket{\Psi_0}$, and $\ket{\Psi(t)}$ satisfies the dynamical equation (\ref{eq:SE}),
or (b) $\op{\rho}(t_F)=\op{\rho}_1$, $\op{\rho}(t_0)=\op{\rho}_0$, and $\op{\rho}(t)$ 
satisfies the dynamical equation (\ref{eq:LE}) [or (\ref{eq:LE2}) in the general 
case], where $\op{\rho}_0$ [$\ket{\Psi_0}$] and $\op{\rho}_1$ [$\ket{\Psi_0}$]
represent the initial and target state, respectively.

The final problem, optimization of observables, requires finding a Hamiltonian
such that the expectation value (or ensemble average) of an observable $\op{A}$,
$\Tr[\op{A}\op{\rho}(t_F)]$, assumes a maximum or minimum at a certain time $t_F$, 
given that the state $\op{\rho}(t)$ evolves according to Eq.~(\ref{eq:LE}) and 
satisfies an initial condition $\op{\rho}(t_0)=\op{\rho}_0$.  Problems of this 
type arise frequently in atomic and molecular physics and chemistry, where the 
observables of interest can range from the position or momentum of a particle, 
to the dipole moment or the vibrational energy of a molecular bond, etc., but 
they are also relevant in quantum computing where we wish to maximize the gate 
fidelity or the projection of the system onto subspaces that are robust with 
regard to decoherence, etc.

While the second problem may appear much simpler than the first, it can be shown 
that for a \emph{generic quantum ensemble}, i.e., a quantum ensemble described by
a density operator $\op{\rho}$ with a maximum number of distinct eigenvalues, the 
problems of quantum state and process engineering are essentially equivalent, up 
to a usually unobservable (and hence insignificant) global phase factor. Similarly, 
the third problem is equivalent to quantum process engineering if the initial state
of the system is a generic quantum ensemble and the target observable is represented 
by an operator $\op{A}$ with distinct eigenvalues (occurring with multiplicity 1). 
It should be noted, however, that this equivalence does not hold for pure-state 
systems, which are always represented by density operators of rank 1, and for 
which process engineering is in general a much harder problem than the others. 
[See e.g.~\cite{JPA35p8315}]

\section{Hamiltonian Engineering}

Having shown how many problems in quantum control can be reduced to Hamiltonian
engineering problems, we shall now consider various strategies for finding and 
implementing control Hamiltonians for quantum systems, and their advantages and 
drawbacks.  So far, we have not made any assumptions about the structure of the 
Hamiltonian $\op{H}[\vec{f}(t)]$ or the nature of the control fields $\vec{f}(t)$ 
applied, both of which depend on the specific physical systems considered.  While
we will try to avoid being too specific some additional assumptions are necessary.

We can always partition $\op{H}[\vec{f}(t)]$ into a system part $\op{H}_S$, which 
describes the system's intrinsic dynamics and is independent of the controller, 
and a control part $\op{H}_C[\vec{f}(t)]$.  Although $\op{H}_C[\vec{f}(t)]$ can 
depend on the control functions $f_m(t)$ in a nonlinear fashion, in many situations
assuming a linear dependence on the field components $f_m(t)$
\begin{equation} \label{eq:CL}
  \op{H}_C[\vec{f}(t)] = \sum_{m=1}^M f_m(t) \op{H}_m,
\end{equation}
\vspace{-2ex}
is a reasonably good approximation.  Furthermore, while in some applications such
as solid-state architectures with multiple control electrodes, there are naturally 
multiple independent fields, in many applications there is only a single effective 
control field such as the electromagnetic field induced by a laser pulse, a maser 
or radio-frequency field, for instance, and thus the control Hamiltonian simplifies 
further $\op{H}_C[\vec{f}(t)]=f(t)\op{H}_1$.  

In the control-linear case~(\ref{eq:CL}), a necessary and sufficient condition for
being able to engineer any (unitary) process up to a global phase factor is that 
the Lie algebra generated by $i\op{H}_S$ and $i\op{H}_m$, $m=1,\ldots,M$ is either 
$u(N)$ or $su(N)$, where $N$ is the dimension of the relevant Hilbert space $\H$
[\cite{qph0106128,JPA35p4125}].  As outlined in the previous section, this is also
a necessary and sufficient condition for quantum state or observable controllability, 
at least for generic quantum ensembles and observables.  Most quantum systems can 
be shown to be controllable, but constructive control can be challenging.  

\subsection{Geometric Control Techniques}

Given a set of (independent) control Hamiltonians $\op{H}_m$, $m=1,\ldots,M$, which
is complete in that the $i\op{H}_m$ generate the entire Lie algebra, the simplest 
general strategy we can pursue is to expand the (unitary) target process $\op{U}$ 
into a product of elementary (complex) rotations $\exp(ic\op{H}_m)$ in a cyclic, 
iterating pattern
\begin{equation} \label{eq:prod}
  \op{U} = \prod_{k=1}^K \left[ \prod_{m=1}^M \exp(i c_{km} \op{H}_m) \right]
\end{equation}
\vspace{-2ex}
and determine the constants $c_{km}$ in the expansion using Lie group decomposition
methods, from which we can derive suitable values for the field strengths $f_{mk}$ 
and control pulse lengths $t_{mk}$ via the relation $c_{mk}=\int_0^{t_{mk}} f_{mk}(t) 
\,dt$, which reduces to $c_{mk}=f_{mk}t_{mk}$ for piecewise constant fields.  

If the $\op{H}_m$ are orthogonal, i.e., $\Tr[\op{H}_m\op{H}_n]=\mbox{const. }
\delta_{mn}$, there are well-developed geometric techniques to solve this type of 
problem.  Explicit solutions have been developed especially for many problems 
involving spin$-1/2$ particles, which are of interest in nuclear magnetic resonance 
(NMR) applications 
[e.g.~\cite{IEEE39CDC1074}], and even constructive algorithms for the generation of
arbitrary unitary operators on $n$-qubits via $SU(2^n)$ decompositions have been 
proposed recently [\cite{JMP46p001}].  For non-qubit systems the problem is harder 
but some explicit decomposition algorithms for simple $N$-level systems have also 
been proposed [e.g.~\cite{PRA61n032106}].  

In practice, however, most physical systems are subject to internal system dynamics
in addition to the control-induced dynamics.  This leads to a drift term $\op{H}_S$,
which is usually non-trivial and cannot be turned off.  In principle, we can include 
this term in the expansion~(\ref{eq:prod}) by replacing the control Hamiltonians 
$\op{H}_m$ by $\op{H}_S+\op{H}_m$, for example.  However, even if the $\op{H}_m$ are 
orthogonal, the new effective Hamiltonians in the expansion are usually not, which 
significantly complicates the problem.  Moreover, even if we find a decomposition 
for given a problem, it may not be physically realizable as the $c_{km}$ could be 
negative.  Since the control coefficient corresponding to $\op{H}_S$ is fixed 
($f_S=1$), implementing such a rotation would require letting the system evolve for 
negative times, which is usually impossible (except possibly if the evolution of the
system is periodic).  Decomposition techniques for non-orthogonal Hamiltonians that 
take these constraints into account exist for $SU(2)$ [\cite{PRA62n053409}] but even
in this simple case the resulting pulse sequences are much more complicated and for 
higher-dimensional systems even the minimum number of pulses necessary to generate a
given unitary transformation given non-orthogonal Hamiltonians is generally not known.
  
The usual way to circumvent the problem of drift is by transforming to a rotating 
frame (RF) given by $e^{-i t E_n} \ket{n}$, where $\{\ket{n}: n=1,\ldots, N=\dim \H\}$ 
is a basis of $\H$ consisting of eigenstates  $\ket{n}$ of $\op{H}_S$ with eigenvalues 
(energies) $E_n$, $\op{H}_S\ket{n}=E_n\ket{n}$.  Setting $U_S(t)=\exp(-it\op{H}_S)$ 
the dynamics in the rotating frame is governed by the new (interaction picture) 
Hamiltonian
\begin{equation}
  \op{H}_C'[\vec{f}(t)] = \op{U}_S(t)^\dagger \op{H}_C[\vec{f}(t)] \op{U}_S(t).
\end{equation}
\vspace{-2ex}
Thus we have transformed away the drift term but the previously time-independent 
Hamiltonians $\op{H}_m$ in the control-linear approximation are now time-dependent
\begin{equation}
  \op{H}_m'[\vec{f}(t)] = \op{U}_S(t)^\dagger \op{H}_m \op{U}_S(t).
\end{equation}
\vspace{-2ex}
Thus further approximations such as spectral decomposition of the control fields 
into a small number of distinct frequency components are needed, e.g.,
\begin{equation} \label{eq:f}
  f(t) = \sum_{n,n'>n} A_{nn'}(t) \cos(\omega_{nn'}t + \phi_{nn'}),
\end{equation}
\vspace{-2ex}
where $\omega_{nn'}=(E_{n'}-E_n)/\hbar$ is the transition frequency between states 
$\ket{n'}$ and $\ket{n}$, $A_{nn'}(t)$ are ``amplitude functions'' and $\phi_{nn'}$ 
constant phases.  If the system is strongly regular, $\omega_{nn'}\neq\omega_{mm'}$ 
unless $(n,n')=(m,m')$, and the transition frequencies $\omega_{nn'}/\hbar$ are 
well separated, then choosing control fields of the form~(\ref{eq:f}) with slowly 
varying amplitude functions $A_{nn'}(t)$ compared to $2\pi/\omega_{nn'}$, enables 
us to address individual transitions via frequency selective pulses, and assume 
the following decomposition of the control Hamiltonian
\begin{equation}
  \op{H}_C' = \sum_{n,n'>n} f_{nn'}(t) \op{U}_S(t)^\dagger \op{H}_{nn'} \op{U}_S(t)
\end{equation}
\vspace{-2ex}
where $f_{nn'}(t)=A_{nn'}(t)\cos(\omega_{nn'}t+\phi_{nn'})$.  We can further decompose 
the fields into rotating and counter-rotating terms since $2\cos(\omega_{nn'}t+\phi) 
= e^{+i(\omega_{nn'}t+\phi)}$ $+e^{-i(\omega_{nn'}t+\phi)}$.  Noting that $\op{U}_S(t)$ is 
diagonal with respect to the eigenbasis $\ket{n}$ with elements $e^{-it E_n}$, and 
making the rotating wave approximation (RWA), i.e., assuming that the pulses are 
\emph{sufficiently long} that the contribution of the counter-rotating terms
averages to zero, we obtain the simplified RWA control Hamiltonian
\begin{equation} \label{eq:HRWA}
  \op{H}_C^{RWA} = \sum_{n,n'>n} \Omega_{nn'}(t)\op{H}_{nn'}(\phi_{nn'})  
\end{equation}
\vspace{-2ex}
where $\Omega_{nn'}=A_{nn'}(t) d_{nn'}/2\hbar$, and $\op{H}_{nn'}(\phi_{nn'})= 
\op{x}_{nn'}\cos\phi_{nn'}+\op{y}_{nn'}\sin\phi_{nn'}$, $d_{nn'}$ being the transition 
dipole moment and 
   $\op{x}_{nn'} = \ket{n}\bra{n'}+\ket{n'}\bra{n}$.  
   $\op{y}_{nn'} = i(\ket{n}\bra{n'}-\ket{n'}\bra{n})$.
Thus, the evolution of the system is now governed by a control Hamiltonian that is 
drift-free with time-independent components $\op{H}_{nn'}$, as desired.  

The RF and RWA are ubiquitous in physics, and the RWA control Hamiltonian is the 
starting point for the design of control schemes using geometric techniques in 
many applications.  As the derivation showed, however, it involves several 
simplifying assumptions such as strong regularity of the system, control dynamics
much slower than the intrinsic dynamics of the system (control pulses must be much
longer than the oscillation periods $2\pi/\omega_{nn'}$ for (\ref{eq:f}) and RWA to 
make sense), and the negligibility of off-resonant excitation.  In reality, there 
are often further complications.  E.g., for coupled spin systems one typically 
transforms into a multiply-rotating frame which leaves a residual drift term, 
which is usually neglected by assuming that the control pulses are sufficiently 
strong (hard) so that the influence of the drift term is negligible and the 
Hamiltonians can be assumed to be effectively orthogonal for the decomposition.  
Obviously, these assumptions are not always valid.  [For an analysis of the 
validity of geometric control schemes for atomic and molecular systems see 
e.g.~\cite{JPA35p8315}.] 

\subsection{Optimal Strategies for Fast Control}

If we wish to achieve control on time-scales comparable to the intrinsic dynamics
of the system, or wish to control systems with too many or insufficiently distinct 
transition frequencies, as is the case for complex molecular systems in chemistry
for example, then a different approach is required.  A promising alternative is to
formulate the control problem in terms of optimization of an objective functional 
such as the distance from a target state or target process, the expectation value 
of an observable or the gate fidelity, subject to the constraint that the dynamical
evolution equations~(\ref{eq:SE}), (\ref{eq:LE}) or (\ref{eq:LE2}) be satisfied.  

A particularly flexible approach is to formulate the control problem 
as a variational problem by defining a functional $\J=\A-\D-\C$, which incorporates 
the objective $\A$, the dynamical constraints $\D$ and the costs $\C$ and depends 
on variational trial functions, and to use variational calculus to find necessary
and sufficient conditions for an extremum.  For instance, a very popular choice in
physical chemistry involves setting $\A=\Tr[\op{A}\op{\rho}(t_F)]$, where $\op{A}$ 
is a (Hermitian) operator representing the objective (e.g., $\op{A}$ could be the
projector $\ket{\Psi}\bra{\Psi}$ onto a target state $\ket{\Psi}$), with a dynamical 
constraint functional given by
\begin{equation}
   \D = \int_{t_0}^{t_F} 
   \Tr\left[A_v(t) \left(\dot{\rho}_v+ \L[\rho_v(t)] \right)\right] \, dt
\end{equation}
\vspace{-2ex}
where $\L[\rho_v(t)]=\L_H[\rho_v(t)]+\L_M[\rho_v(t)]+\L_D[\rho_v(t)]$ and $\rho_v(t)$ 
and $A_v(t)$ are variational trial functions for the state (density operator) and 
observable, respectively, and a cost term related to the control field energies 
\begin{equation}
\vspace{-2ex}
  \C=\sum_{m=1}^M \frac{\lambda_m}{2\hbar} \int_{t_0}^{t_F} |f_m(t)|^2
\end{equation}
\vspace{-2ex}
where $\lambda_m$ are penalty weights. Setting the independent variations of $\J$ 
with regard to $A_v$, $\rho_v$ and $f_m$ to 0, a necessary condition for $\J$ to 
have an extremum, then leads to a set of coupled differential equations with mixed
boundary conditions, the Euler-Langrange equations, which can be solved numerically
to obtain a solution for the control fields $f_m(t)$ and the corresponding
trajectories for the state $\rho(t)$ and the observable $A(t)$.

The key advantages of this approach are that we can deal with complex (and even 
non-linear) control Hamiltonians (no RWA or other approximations required), and 
are in fact not limited to Hamiltonian systems at all, and a wide range of costs 
or trade-offs can be taken into account.  A drawback is that the Euler-Lagrange 
equations are almost always nontrivial and can only be solved using numerical
techniques.  However, efficient algorithms with good convergence properties exist 
for a large class of problems [see e.g.~\cite{JCP118p8191}], and promising results
have been obtained for various applications, especially laser control of molecular
systems using ultra-fast (sub-picosecond) pulses, a regime well outside the realm 
of applicability for frequency-selective geometric control schemes.

\subsection{Robust Control through Adiabatic Passage}

On the opposite end of the spectrum are adiabatic techniques.  Rather than applying
control fields to induce (fast) transitions between various states of the system by
absorption or emission of field quanta (usually photons), adiabatic techniques rely
on (slow) continuous deformation of the energy surfaces by strong, slowly varying
control fields, and adiabatic following of the system's state, making use of the
eigenstate decomposition of the Hamiltonian 
\begin{equation} \label{eq:ED}
  \op{H}[\vec{f}(t)] = \sum_{n=1}^N \eps_n(t) \ket{\Psi_n(t)}\bra{\Psi_n(t)}.
\end{equation}
\vspace{-2ex}
For a Hamiltonian with a control-induced time-dependence the eigenvalues $\eps_n(t)$
and corresponding eigenstates $\ket{\Psi_n(t)}$ vary in time.  The idea of adiabatic
passage is that if we start in a particular initial state $\ket{\Psi_0}$, which will
usually be an eigenstate of the system's intrinsic Hamiltonian $\op{H}_S$, and then
\emph{slowly} switch on and vary suitable control fields $\vec{f}(t)$, then rather 
than inducing spontaneous transitions to other states, the state $\ket{\Psi(t)}$ is
going to remain an eigenstate of the Hamiltonian and follow the path determined by 
the control fields via Eq.~(\ref{eq:ED}).  Adiabatic passage is the basis for many 
control schemes in atomic physics [\cite{90Shore}], most notably STIRAP.

\begin{figure}
\center\myincludegraphics[height=1.5in]{figures/png/Lambda.pdf}{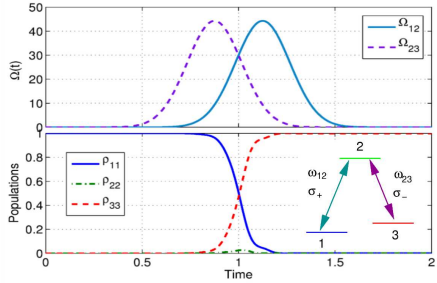}
\caption{STIRAP pulse sequence (top) for robust population transfer from state 
$\ket{1}$ to state $\ket{3}$ (bottom) for three-level $\Lambda$ system (inset).} 
\label{fig2}
\end{figure}

The simplest example is population transfer in a three-level $\Lambda$-system (see 
Fig.~\ref{fig2}) simultaneously driven by two fields $f_m(t)= A_m(t)\cos(\omega_m t)$, 
$m=1,2$, that resonantly exite the $1\rightarrow 2$ and $2\rightarrow 3$ transitions, 
respectively.  In this case the RWA Hamiltonian~(\ref{eq:HRWA}) simplifies to
\begin{equation}
  \op{H}^{RWA}_C = -\left[ \begin{array}{ccc} 
                   0 & \Omega_1(t) & 0 \\ 
                   \Omega_1(t) & 0 & \Omega_2(t) \\ 0 & \Omega_2(t) & 0 
                  \end{array} \right]
\end{equation}
with $\Omega_m = d_{m,m+1} A_m/2\hbar$ for $m=1,2$.  Setting $\Omega=\sqrt{\Omega_1^2
+\Omega_2^2}$ and $\theta(t)=\arctan[\Omega_1(t)/\Omega_2(t)]$, it is easy to verify 
that the eigenstates of this Hamiltonian are $\ket{\Psi_\pm(t)}=\Omega_1\ket{1}\pm
\Omega\ket{2}+\Omega_2\ket{3}$ for $\lambda_\pm=\pm\Omega$, respectively, and 
$\ket{\Psi_0(t)} = \cos\theta(t)\ket{1}-\sin\theta(t)\ket{3}$ for $\lambda_0=0$.  
Thus, if the system is initially in state $\ket{1}$, then applying control fields 
such that $\Omega_1(t)/\Omega_2(t)$ changes (sufficiently slowly) from $0$ at $t_0$ 
to $\infty$ at $t_F$, results in adiabatic passage of the system from state $\ket{1}
=\ket{\Psi_0(0)}$ to $\ket{3}=\ket{\Psi_0(t_F)}$ as $\theta(t)$ goes from $0$ to 
$\pi/2$.  STIRAP is based on the realization that we can achieve this simply by 
applying two overlapping Gaussian pulses in a \emph{counter-intuitive} sequence, 
i.e, so that $\Omega_2(t)$ starts and ends before $\Omega_1(t)$ as shown in Fig.%
~\ref{fig2}.  Since the upper level $\ket{2}$ is decoupled, i.e., not populated 
during this process, the transfer is robust against decay from the excited state,
which is often a major limiting factor in the control of atomic or molecular 
systems, although it must be noted that the scheme is sensitive with regard to
other forms of decoherence etc.

\subsection{Learning Control via Closed-Loop Experiments}

A potential problem with all of the techniques described so far is their reliance
on a model of the system.  Unfortunately, such models are not always available,
especially for complex systems.  While improved techniques for quantum control 
system identification will hopefully eventually allow us to overcome this problem, 
an alternative which has already been successfully demonstrated in the laboratory
[see e.g.~\cite{PRA63n063412}] are experimental closed-loop learning techniques 
[\cite{PRL68p1500}], which must not be confused with real-time quantum feedback 
control [\cite{PRA49p2133}].  While both schemes rely on (classical) information
gained from measurements, the former involve repeated experiments on many copies
of the system, while the latter rely on continuous observation of the same system
via weak measurements (e.g.\ homodyne detection or passive photodetectors).

Closed-loop learning techniques essentially adaptive, using feedback to guide an 
evolutionary process.  There are many variations but the basic strategy is simple.  
We define 
an objective functional, usually called fitness function, and select initial set 
(population) of control fields.  Each of these is then applied to an identical 
\emph{copy} of the system, and the observable measured to determine the fitness 
of the field.  Then a new generation of control fields is computed from a subset 
of (mostly) well-performing fields using a set of predefined rules for mutations 
and crossovers, and the experiments are repeated with the new generation of fields
until we have arrived at a population of fields with a sufficiently high fitness.  

Although there are some disadvantages to this approach (requirement of many 
identical copies of the system or the ability to efficiently re-initialize the 
system in the same state after each experiment, need for high experimental duty
cycles due to slow convergence, etc.), closed-loop learning techniques have proved 
useful in the laboratory for a wide range of systems, and incorporating feedback 
from such closed-loop learning experiments in some form is likely to be essential 
for quantum control to succeed in the laboratory.  A particularly promising avenue
may be adaptive system identification strategies based on feedback from closed-loop 
experiments.


\vspace{-2ex}
\ack
This work was supported by the Cambridge-MIT Institute's Quantum Technology Project.

\vspace{-2ex}
\bibliography{papers,books,sonia}
\end{document}